\documentclass[conference]{IEEEtran}
\IEEEoverridecommandlockouts
\usepackage{cite}
\usepackage{amsmath,amssymb,amsfonts}
\usepackage{algorithmic}
\usepackage{graphicx}
\usepackage{textcomp}
\usepackage{xcolor}
\def\BibTeX{{\rm B\kern-.05em{\sc i\kern-.025em b}\kern-.08em
    T\kern-.1667em\lower.7ex\hbox{E}\kern-.125emX}}
\begin{document}

\title{A JPEG-based image coding solution \\for data storage on DNA
\thanks{This project has received funding from the European Union's Horizon 2020 research and innovation programme under grant agreement No 863320.}}
\author{\IEEEauthorblockN{ Melpomeni Dimopoulou$^{1}$, Eva Gil San Antonio$^{2}$, Marc Antonini$^{3}$}
\IEEEauthorblockA{\textit{Université Côte d'Azur, CNRS}\\ 
\textit{Laboratoire I3S, UMR 7271 } \\
Sophia Antipolis, France \\
$^{1}$dimopoulou@i3s.unice.fr, $^{2}$gilsanan@i3s.unice.fr, $^{3}$am@i3s.unice.fr}
}

\maketitle

\begin{abstract}
The efficient storage of digital data is becoming very challenging over the years due to the exponential increase in the generation of data which can't compete with the existing storage resources. Furthermore, the infrequently accessed data can be safely stored for no longer than 10-20 years due to the short life-span of conventional storage devices. To this end, recent studies have proven DNA to be a very promising candidate for the long-term storage of digital data. Several pioneering works have proposed different encoding methods for the specific encoding of images into a quaternary DNA representation while first compressing the image using the classical JPEG standard to reduce the high cost of DNA synthesis. However this type of compression is not optimized with respect to the quaternary DNA code and results in an open-loop workflow. In our previous works we have introduced the first closed-loop solution for generating a constrained fixed-length quaternary code which allows controlling the synthesis cost thanks to a source allocation algorithm. In this paper, we extend our studies to proposing a variable-length encoding solution which is based on the JPEG standard.
\end{abstract}


\section{Introduction}
The exponential growth of the digital universe imposes a great challenge in the storage of digital data which is handled by large data centers. An important fraction of this data is accessed infrequently but needs to be safely stored due to security and regulatory compliance reasons. Such data is characterized as cold and is normally stored into off-line back up tape drives which constitute a cheaper means of storage. Nevertheless, conventional storage devices have a limited life-span varying from 10 to 20 years and therefore data needs to be frequently migrated into new storage units, a fact which causes huge energy and hardware waste and is expensive in terms of money. To this end, studies have recently proposed the use of DNA as a novel and promising means of digital data storage which can store 215 petabytes in a single gram and can promise reliable storage for hundreds of years. This last claim can be proven by the successful retrieval and decoding of the DNA of a woolly mammoth that had been trapped into permafrost. DNA is the storage medium of heredity information and consists of 4 different building blocks, the nucleotides (nts) which are denoted using the symbols A, T, C and G. 

DNA coding is a novel solution for the long term storage of digital data which is highly promising, yet extremely challenging. The main goal of this emerging field of study is the encoding of any digital data into a quaternary representation using the four different DNA nucleotides A, T, C, and G. Once a quaternary encoded strand is created it can be written into a DNA strand using a biological process of DNA synthesis which allows the creation of synthetic DNA strands (oligos) which will have the same content as the encoded sequences of digital data. The DNA oligos can then be safely stored into some special capsules that protect the DNA from any contacts with water and oxygen and can guarantee lossless storage for hundreds of years. The stored DNA content can be read from the stored oligos using the biological process of DNA sequencing. While fundamental for DNA data storage, the two biological procedures of DNA synthesis and sequencing impose some difficulties to the encoding of digital data.To begin with, DNA synthesis is an expensive process and costs several dollars per synthesized oligo. Therefore, to reduce this cost, the input data should be efficiently compressed during encoding. Then, sequencing is prone to errors and requires respecting several constraints in the encoding for reducing the probability of noise during decoding. More precisely, the encoding must respect the following rules: 
\begin{itemize}
    \item \textbf{Rule of balanced G,C content:} The encoded strand should not contain a content of C's and G's which is higher than the content of A's and T's 
    \item \textbf{Rule of homopolymers:} There should be no repetitions of the same nucleotide more than 3 consecutive times. 
\end{itemize}
\begin{figure*}[h]
\begin{center}
    \includegraphics[width=0.9\textwidth]{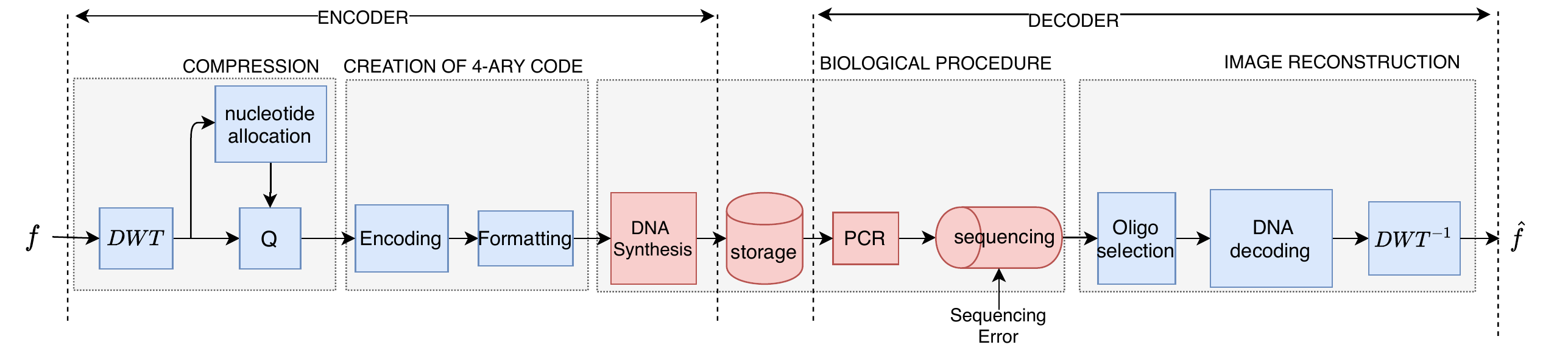} 
\end{center}
\caption{Fixed-length DNA coding workflow introduced in our previous works}
\label{fig:old_workflow}
\end{figure*}
Even though there have been multiple encoding methods introduced by the state of the art (\cite{grass2015robust},\cite{blawat2016forward},\cite{bornholt2016dna},\cite{erlich2016capacity}) which respect the above restrictions, none of the proposed methods provided a simple fixed-length code construction algorithm which would allow implementing a closed-loop encoding solution for controlling the compression and thus the corresponding synthesis cost. Most of the existing works in bibliography have been instead transcoding a JPEG encoded binary stream (\cite{yazdi2015rewritable},\cite{yazdi2017portable},\cite{goldman2013towards}). In our work in \cite{dimopoulou2019biologically}, we proposed such an algorithm, which we have named as PAIRCODE, and evaluated its performance by including it in an end-to-end encoding/decoding workflow which used a source-allocation to optimize the compression performed using a DWT and a Uniform Scalar Quantizer (SQ). In \cite{dimopoulou:hal-02959330}, we have enhanced the performance of the above workflow by replacing the SQ with a Vector Quantizer (VQ). The proposed work-flow performed a fixed-length encoding. In this work we propose a novel variable-length encoder for the storage of digital images into DNA which is inspired by the binary JPEG protocol and is adapted to the quaternary nature of DNA and the needs of DNA data storage. To facilitate understanding, in section \ref{sec:previous_work} we will briefly remind our previous contributions. In section \ref{sec:JPEG-DNA} we present the new variable-length encoding solution and in \ref{sec:results}, we provide the results on the preformance of the proposed algorithm. More precisely, in section \ref{ssec:comp_solutions} we  present the comparison of the different encoding solutions in terms of compression efficiency while in section \ref{ssec:robustness} we test the robustness of the different methods to sequencing noise. Finally, section \ref{sec:conclusions} concludes this work.

\section{Fixed-length encoding using PAIRCODE}
\label{sec:previous_work}

In \cite{dimopoulou2019biologically}, we have introduced the PAIRCODE algorithm for the encoding of digital data into DNA. This code-construction method is fixed-length and in contrast to most of the methods proposed by the state of the art, it can be applied to any type of input data without being restricted to binary inputs. Consequently, it can be easily embedded to any desired encoding workflow while being adapted to the needs of DNA data storage. The PAIRCODE algorithm works according to the following principles.

The design of the constrained quaternary code $\mathcal{C}^*$ is based on the construction of DNA codewords using mainly pair-elements from the following dictionary:
$$\mathcal{C}_1=\{AT,AC,AG,TA,TC,TG,CA,CT,GA,GT\}$$
The concatenation of elements selected from the above dictionary guarantees that the encoding will not contain any homopolymers or a GC-percentage which is higher than the AT-percentage. As the selection of elements from $\mathcal{C}_1$ results in the creation of codewords of an even length, we also allow the creation of odd length codewords by choosing elements from $\mathcal{C}_1$ adding one last element from $\mathcal{C}_2=\{A, T, C, G\}$ at the end of each codeword. The encoding workflow to which the above algorithm has been included is depicted in figure \ref{fig:old_workflow} and works according to the following steps.

The input image is decomposed into different subbands using a DWT.
Each subband is independently quantized using a selected quantizer ($Q$). The quantization is optimized thanks to a source allocation algorithm which aims in finding the quantization parameters that minimize the distortion of each subband for a given target compression rate. The quantized subbands are encoded using the proposed PAIRCODE algorithm to encode each quantized coefficient. The encoded strands are cut into smaller chunks (oligos) of length 200-300 nts and formatted using any necessary headers to denote their order in the long encoded sequence. This formatting process is necessary to ensure an error-free DNA synthesis as the synthesis error increases exponentially for longer DNA strands. The oligos are synthesized in DNA and stored.
When one wants to read back the oligos (decoding step), the synthesized data is amplified using Polymerase Chain Reaction (PCR) which is an enzymatic process that creates multiple copies of DNA strands, adding the redundancy needed for error correction.
Then, the process of sequencing will provide multiple copies, which might contain errors, of each synthesized and amplified oligo. Finally, the consensus sequences are estimated from the noisy copies, which are used to decode and reconstruct the stored data by following the inverse process of the one used in the encoding.

\begin{figure*}[t]
\begin{center}
    \includegraphics[width=0.79\textwidth]{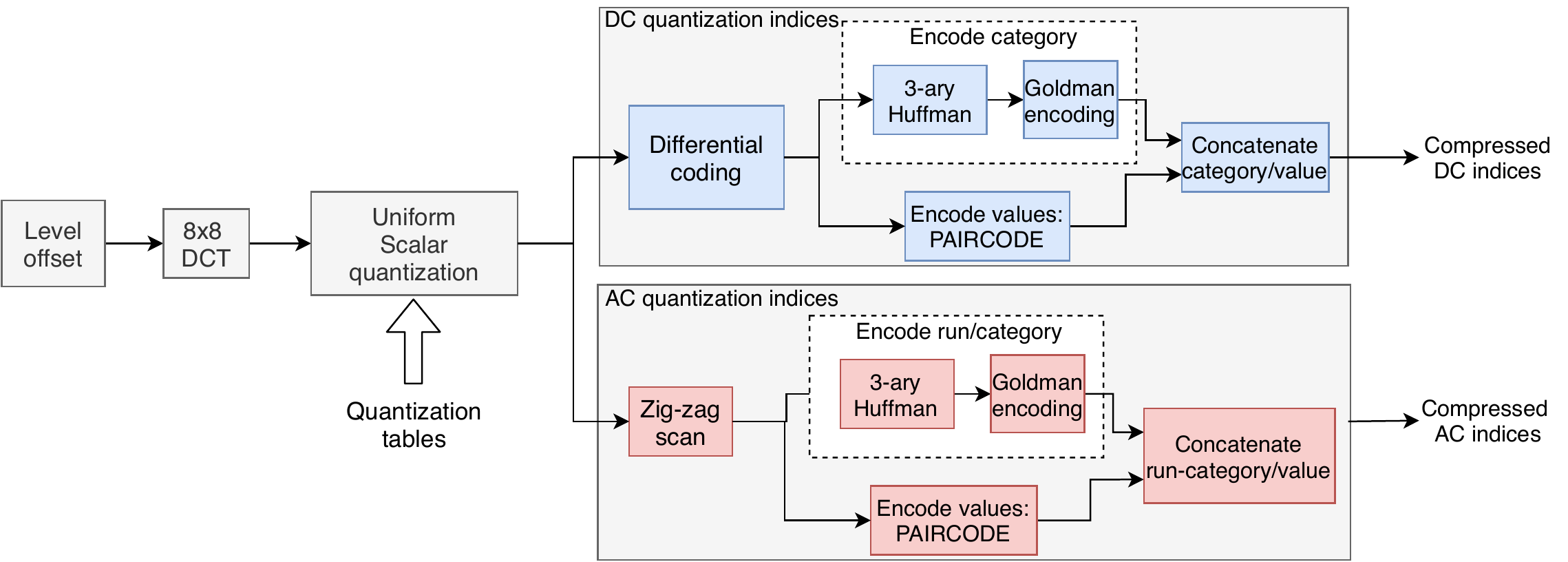} 
\end{center}
\caption{\footnotesize{Workflow of the modified JPEG workflow (JPEG-DNA) to suit the needs of DNA coding}}
\label{fig:jpeg_dna_workflow}
\end{figure*}
 In \cite{dimopoulou2019biologically} we introduced the above workflow for fixed-length encoding and have proposed using a scalar quantizer for the compression. In our latest work in \cite{dimopoulou:hal-02959330}, we have improved the workflow's efficiency by replacing scalar quantizer with a Vector Quantizer (VQ). The obtained results were much more promising as VQ allowed capturing information regarding the distribution of the image but this solution required knowledge of the codebook in the decoder. Consequently, in this new study we propose a new variable-length encoder which is optimizing the compression similarly to the classical JPEG standard. More precisely, in this work we extend the JPEG algorithm to a constrained quaternary representation which is using a combination of our proposed PAIRCODE algorithm and the encoding algorithm proposed by Goldman \textit{et al.} \cite{goldman2013towards}.

\section{JPEG-DNA: A modified JPEG coder for DNA coding}
\label{sec:JPEG-DNA}
Inspired by the main workflow of the classical JPEG standard, in this work we propose JPEG-DNA, a modified version of the JPEG algorithm for the encoding of an image into a constrained quaternary representation of A, T, C and G. Classical JPEG provides an optimal encoding expressed into a binary representation by using two main coding techniques. The first technique is the Huffman coding used to encode the run/category of the AC indices produced by a Discrete Cosine Transform, as well as the category of the DC indices. Huffman codes take into consideration the frequency of each run/category AC element or category DC element throughout the full image and assigns binary words to the different values while ensuring that each binary representation is not a prefix of another. The main asset of Huffman coding is that the algorithm assigns the shortest words to the most frequent elements so to enhance the performance of the encoding in terms of compression. Then, the second method used in JPEG is a simple binary coding of the values of AC and DC indices. This encoding is simply transforming each value into its binary representation using a number of bits which is predefined by the category field that precedes. It is therefore clear, that to modify the existing JPEG algorithm so to provide a quaternary representation one would need to replace those two binary encodings by some appropriate quaternary ones while respecting the encoding constraints of DNA coding.

In the works of Goldman et al. in \cite{goldman2013towards}, the authors presented an interesting algorithm for encoding a sequence of symbols to a constrained DNA sequence. One of the main assets of this algorithm compared to other state of the art methods is that similarly to our proposed constrained codebook it can be applied to any type of input without being restricted to binary inputs. The algorithm proposed by Goldman encodes the input data using a ternary Huffman into a stream of the trits 0, 1 and 2. Then, trit is replaced with one of the nucleotides, excuding the one which was previously used, this way ensuring that no homopolymers are generated. Hence, since the above algorithm is using Huffman codes it seems to be an interesting candidate to replace the binary Huffman encoding of the classical JPEG standard.

\begin{table}[htb]
\centering
\begin{minipage}[b]{0.8\linewidth}
\resizebox{\textwidth}{!}
{
\begin{tabular}{|c|c|}
\hline
 Range & \begin{tabular}{@{}c@{}}Category \\ (\#nt used for encoding) \end{tabular} \\ \hline
 0 & 0   \\ 
 -5, $\hdots$ , -1,1, $\hdots$ , 5 & 2   \\ 
 -25,$\hdots$ , -6,6, $\hdots$ , 25  & 3 \\ 
 -75,$\hdots$ , -26,26, $\hdots$ , 75 & 4 \\ 
 -275,$\hdots$ , -76,76, $\hdots$ , 275 & 5 \\ 
 -775,$\hdots$ , -276,276, $\hdots$ , 775 & 6 \\ 
 -2775,$\hdots$ , -776,776, $\hdots$ , 2775 & 7 \\
 -7775,$\hdots$ , -2776,2776, $\hdots$ , 7775 & 8 \\ \hline
\end{tabular}
}
\caption{Category range for encoding values using DNA. Category 1 is omitted due to the biological constraints imposed by DNA sequencing.}
\label{table:jpeg_dna_categories}
\end{minipage}
\end{table}

\begin{figure}[t]
\centerline{
\includegraphics[scale=0.6]{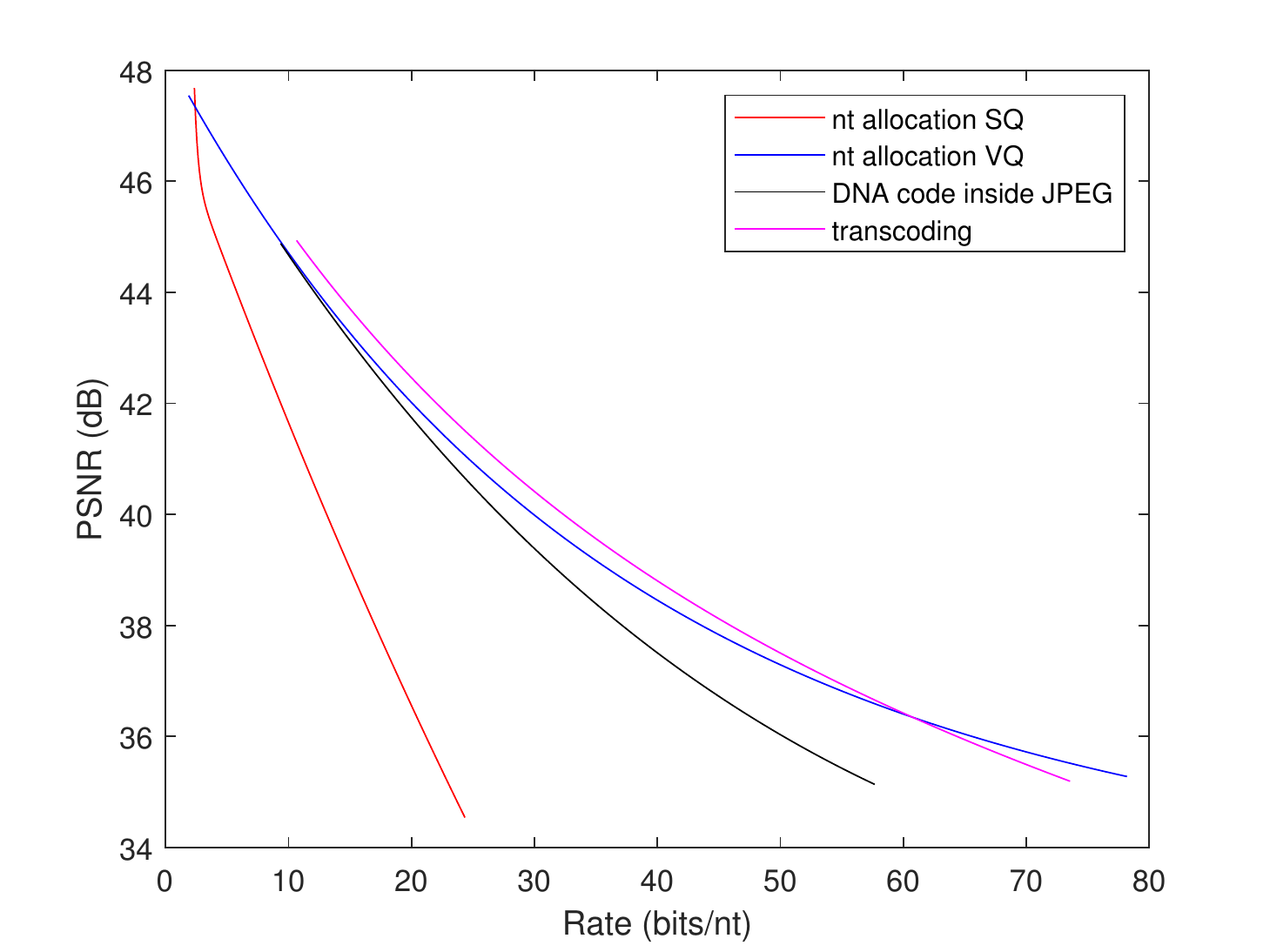} 
}
\caption{Comparison of the different proposed encoding solutions. The methods compared are the following: nucleotide allocation using SQ and PAIRCODE for the encoding (red), nucleotide allocation using VQ and PAIRCODE for the encoding (blue), JPEG transcoding (magenta) and JPEG with the quaternary code included in the optimization (black).}
\label{fig:comparison}
\end{figure}

\begin{figure*}[t]
\centering
	\begin{minipage}[b]{0.22\linewidth}
		\centerline{
           \includegraphics[scale=0.27]{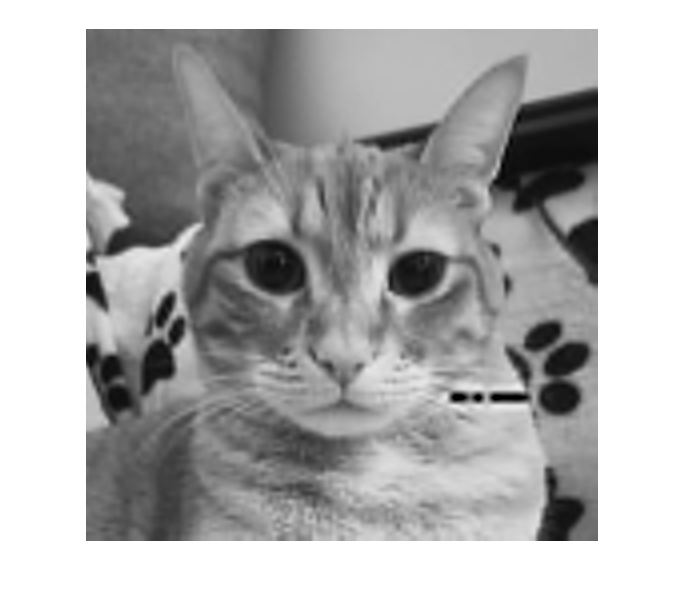}}\vspace{-4.5mm}
        \centerline{
		\begin{tabular}{c}
            \footnotesize{nt allocation using SQ}\\
            \footnotesize{1 deletion in LL subband}	
		\end{tabular}}
	   \end{minipage}
	\begin{minipage}[b]{0.22\linewidth}
		\centerline{
           \includegraphics[scale=0.203]{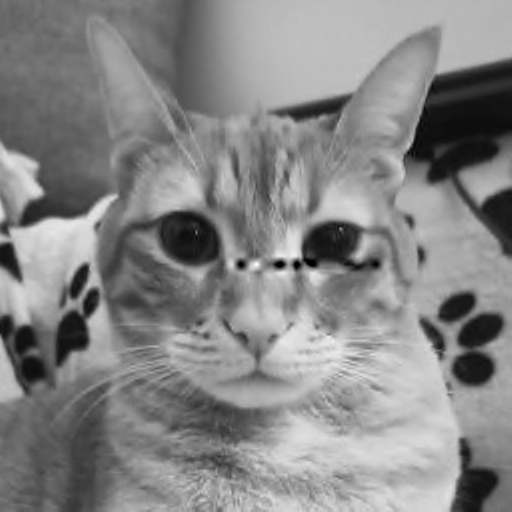}}
        \centerline{
		\begin{tabular}{c}
            \footnotesize{nt allocation using VQ}\\
            \footnotesize{1 deletion in LL subband}	
		\end{tabular}}
	   \end{minipage}
		\begin{minipage}[b]{0.22\linewidth}
		\centerline{
          \includegraphics[scale=0.27]{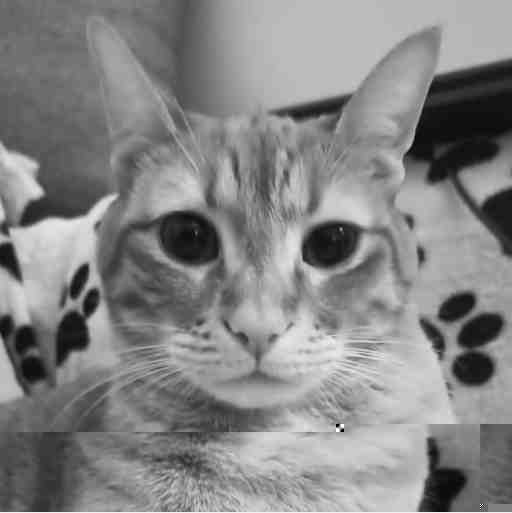}}
        \centerline{
		\begin{tabular}{c}
           \footnotesize{JPEG-DNA}\\
           \footnotesize{1 deletion}
		\end{tabular}}
	   \end{minipage}
		\begin{minipage}[b]{0.22\linewidth}
				\centerline{
           \includegraphics[scale=0.27]{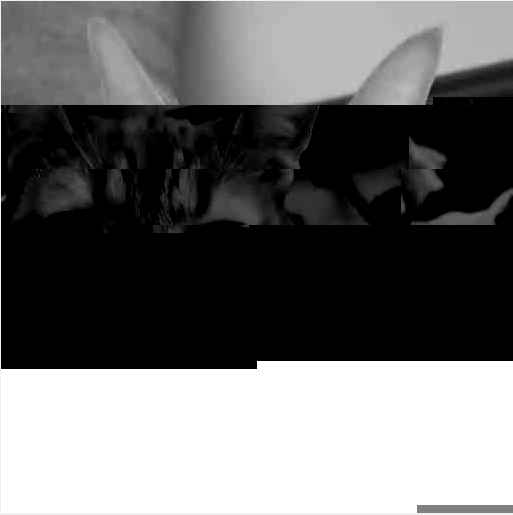}}
      \centerline{
		\begin{tabular}{c}
          \footnotesize{transcoding}\\
          \footnotesize{1 deletion}
		\end{tabular}}
	   \end{minipage}
\caption{The impact of one deletion error on a 512x512 pixel image of a cat for the different encoding solutions. The original images have been compressed to obtain the same PSNR value of 38.5 dB.} 
\label{fig:error_impact_same PSNR}
\end{figure*}
Then, for replacing the binary encoding of JPEG for the non-zero values of the AC and DC indices, we propose using our PAIRCODE algorithm which works similarly to the classical fixed-length binary encoding and provides a quaternary representation while respecting the sequencing constraints. The use of this code will change the range of the categories which will now determine the number of nucleotides rather than the number of bits that will be used for the encoding of an element. Since in our encoding a representation of 1nt per element is not permitted, category 1 is omitted from the list of possible categories as shown in table \ref{table:jpeg_dna_categories}. In other words, in the proposed variable-length algorithm we have used the main workflow of JPEG by replacing the binary Variable Length Coding (VLC), which uses binary Huffman and classical binary encoding, with a quaternary VLC algorithm that uses the algorithm proposed in Goldman et al. and our constrained code (PAIRCODE). More precisely, the run/category of AC indices and the category of DC indices are encoded using a ternary Huffman to produce a sequence of trits, each of which is then encoded to one nucleotide by avoiding the symbol which had been used for encoding the previous position. The values of AC and DC indices are then encoded to some codeword of length $l$ that belongs to a code which has been generated using PAIRCODE. The length $l$ depends on the category to which the value belongs (table \ref{table:jpeg_dna_categories}). The modified workflow is presented in figure \ref{fig:jpeg_dna_workflow}. In the following section we provide experimental results on the performance of this modified JPEG code compared to the fixed length encoding which has been proposed in our previous works. We also provide the results of the transcoding method which encodes the binary output of a classical JPEG using a quaternary fixed length coding for encoding each byte of the JPEG binary stream.

\section{Results}
\label{sec:results}

\subsection{Comparison of the different encoding solutions}
\label{ssec:comp_solutions}
In this section, we compare all the encoding methods proposed in our studies. Namely, we will compare the two fixed-length encoding solutions using a Uniform Scalar Quantizer, as proposed in \cite{dimopoulou2019biologically}, and a VQ as proposed in \cite{dimopoulou:hal-02959330}, to our proposed JPEG-DNA solution and the simpler transcoding method. For the comparison we built the curve of PSNR in function of the coding potential in bits/nt and the result is given in figure \ref{fig:comparison}. As observed, the fixed length encoding with the Scalar Quantizer has the less efficient performance. This result is justified by the fact that, as expected, fixed length solutions are less performant than variable length encodings and in the case of a Scalar quantization the encoder is not taking into consideration the distribution of the source and thus the compression is less adapted to the characteristics of the input image. However, this has been only our very first attempt to build a simple fixed-length encoder which allows controlling the compression rate to test the performance of our proposed PAIRCODE algorithm. 

Interestingly enough, the fixed-length solution which is using a VQ for the quantization, provided results which are comparable to the ones of the variable-length solutions. This improvement in the performance is explained by the fact that in the case of a VQ the encoder uses some knowledge of the characteristics of the input. This information stems from the use of a well-selected training set of images, with similar characteristics to the input image, for the creation of a good codebook of quantization vectors.  

We also observe that among the variable length solutions, simple transcoding shows the best results. In other words, even if the "closed-loop" solution of inserting PAIRCODE into the code of JPEG to allow controlling the compression was expected to perform better, the result is different than predicted. The difference in the performance between the two variable length encoding methods is explained by the constraints imposed on the quaternary encoding. More specifically, in the proposed "closed-loop" solution the category 1 that is using 1 nt to encode a value is omitted in order to avoid the creation of homopolymers. However, in most cases the values of category 1 are the most frequent ones, and thus in the transcoding case those values will be encoded by 1 bit while in the quaternary case they are encoded by 2 nts. This explanation becomes even more apparent when checking how the difference in both cases' performance evolves with the increase of the encoding rate. The more the encoding rate increases, the more frequent values which are found around zero will be quantized to category 1 and thus transcoding will take the lead in terms of performance. Nevertheless, since this has been only a first proposal for implementing an efficient variable-length quaternary encoding for DNA data storage, the proposed algorithm can be further improved to deal with this issue and provide better results.

\subsection{Robustness of the different encoding solutions}
\label{ssec:robustness}
Even though the above experiment is testing the strength of the encoding scenarios in terms of compression quality, one must not neglect the main obstacle of DNA coding which is the error-prone DNA sequencing. It is clear, that even though a fixed length encoding might be less efficient in compression, it is more robust to sequencing noise. This is due to the fact that in the case of an error during sequencing, only a part of the decoding will be affected. To facilitate understanding, let's take the following example. Assuming the case of a deletion error, the full structure of an oligo is affected by shifting all the nucleotides following the deleted-one, by one position. However, since the encoding is fixed-length, the decoding of one oligo does not depend on the decoded information of the previous oligos. Thus, even is one oligo is lost, the following oligo can be correctly decoded. on the other hand, in a variable-length encoding, a deletion error in some oligo will also affect the rest of the decoding and the structure of the image can be lost.

To prove this claim we have tested the decoding of an image in the presence of one single deletion at a random position for the cases of a simple fixed-length encoding using Vector Quantization and the two variable length solutions of transcoding and modified JPEG for DNA. The impact of such an error in the visual quality of the decoded image is presented in figure \ref{fig:error_impact_same PSNR}. As observed, the impact of one deletion error in the fixed-length encoding using VQ affects less the visual quality of the input image and causes the loss of a single vector. However, in the case of a variable length encoding, one deletion will completely change the image reconstruction causing a higher visual distortion. More precisely, in the case of transcoding, the error has a much worse impact on the decoding of the image. 

To give an explanation to this fact, let's consider a pool of oligos which have been selected as the most representative ones after computing the consensus sequences. Then let's assume that all the selected oligos are correct except for one, which has suffered a deletion error. In the case of the JPEG-DNA, one deletion will cost a shift in all the nucleotides that follow the deletion. Therefore, the total impact on the long reconstructed strand will be a shift of all nucleotides following the deleted nucleotide by one position to the left. The same type of error in the case of transcoding, one single deletion will create a shift on the DNA strand after the deleted nucleotide that will cause a wrong decoding of all the bytes that follow the deletion. This creates much more distortion that leads in losing big part of the input image.
It is obvious that, as predicted, the fixed length encoding solutions are much more robust to noise while in the case of variable-length one error can result in losing the structure of the image. It is therefore clear that in DNA coding the selection of the best encoding solution does not only rely upon the performance, but highly depends on the robustness to sequencing error. 

\section{Conclusions}
\label{sec:conclusions}
In this work, we have proposed JPEG-DNA, a DNA image coding solution based on the classical JPEG standard. This solution introduces in the standard codec a variable-length DNA closed-loop coding. Even though the comparison of the different methods presented shows a higher performance when transcoding the binary output of a classical JPEG, the fixed length encoding solutions are much more robust to noise as in the case of variable-length one error can result in losing the structure of the image. Furthermore, among the variable-length solutions, JPEG-DNA is more robust than transcoding. It is therefore clear that in DNA coding the selection of the best encoding solution does not only rely upon the performance, but highly depends on the robustness to errors. 

\bibliographystyle{IEEEbib}
\bibliography{refs}

\begin{thebibliography}{1}

\bibitem{grass2015robust}
Robert~N Grass, Reinhard Heckel, Michela Puddu, Daniela Paunescu, and
  Wendelin~J Stark,
\newblock ``Robust chemical preservation of digital information on {DNA} in
  silica with error-correcting codes,''
\newblock {\em Angewandte Chemie International Edition}, vol. 54, no. 8, pp.
  2552--2555, 2015.

\bibitem{blawat2016forward}
Meinolf Blawat, Klaus Gaedke, Ingo Huetter, Xiao-Ming Chen, Brian Turczyk,
  Samuel Inverso, Benjamin~W Pruitt, and George~M Church,
\newblock ``Forward error correction for {DNA} data storage,''
\newblock {\em Procedia Computer Science}, vol. 80, pp. 1011--1022, 2016.

\bibitem{bornholt2016dna}
James Bornholt, Randolph Lopez, Douglas~M Carmean, Luis Ceze, Georg Seelig, and
  Karin Strauss,
\newblock ``A {DNA}-based archival storage system,''
\newblock {\em ACM SIGOPS Operating Systems Review}, vol. 50, no. 2, pp.
  637--649, 2016.

\bibitem{erlich2016capacity}
Yaniv Erlich and Dina Zielinski,
\newblock ``Capacity-approaching {DNA} storage,''
\newblock {\em bioRxiv}, p. 074237, 2016.

\bibitem{yazdi2015rewritable}
SM~Hossein~Tabatabaei Yazdi, Yongbo Yuan, Jian Ma, Huimin Zhao, and Olgica
  Milenkovic,
\newblock ``A rewritable, random-access {DNA}-based storage system,''
\newblock {\em Scientific reports}, vol. 5, pp. 14138, 2015.

\bibitem{yazdi2017portable}
SM~Hossein~Tabatabaei Yazdi, Ryan Gabrys, and Olgica Milenkovic,
\newblock ``Portable and error-free dna-based data storage,''
\newblock {\em Scientific reports}, vol. 7, no. 1, pp. 1--6, 2017.

\bibitem{goldman2013towards}
Nick Goldman, Paul Bertone, Siyuan Chen, Christophe Dessimoz, Emily~M LeProust,
  Botond Sipos, and Ewan Birney,
\newblock ``Towards practical, high-capacity, low-maintenance information
  storage in synthesized {DNA},''
\newblock {\em Nature}, vol. 494, no. 7435, pp. 77, 2013.

\bibitem{dimopoulou2019biologically}
Melpomeni Dimopoulou, Marc Antonini, Pascal Barbry, and Raja Appuswamy,
\newblock ``A biologically constrained encoding solution for long-term storage
  of images onto synthetic {DNA},''
\newblock in {\em EUSIPCO 2019}, Galicia, Spain, Sept. 2019.

\bibitem{dimopoulou:hal-02959330}
Melpomeni Dimopoulou and Marc Antonini,
\newblock ``{Image storage in DNA using Vector Quantization},''
\newblock in {\em {EUSIPCO 2020}}, Amsterdam, Netherlands, Jan. 2021.

\end{thebibliography}

\end{document}